\documentclass[prb,twocolumn,showpacs,amsmath,amssymb]{revtex4}
\usepackage{graphicx}
\usepackage{graphics}
\usepackage{dcolumn}
\usepackage{bm}
\usepackage{amssymb}
\usepackage{epsfig}
\include{commands.tex}
\newcommand {\el} {\\ \nonumber}

\newcommand {\ket} [1] {| #1 \rangle}
\newcommand {\bkt} [1] {\langle #1 \rangle}
\newcommand {\dbkt} [2] {\langle #1 | #2 \rangle}
\newcommand {\tbkt} [3] {\langle #1 | #2 | #3 \rangle}
\newcommand {\pd} [2] {\frac{\partial #1}{\partial #2}}
\newcommand {\td} [2] {\frac{d #1}{d #2}}

 \newcommand {\beq}{\begin{equation}}
\newcommand {\eeq}{\end{equation}}
\newcommand {\bea}{\begin{eqnarray}}
\newcommand {\eea}{\end{eqnarray}}
\begin{document}
\title{Electrical generation of spin in crystals with reduced symmetry}
\author{Dimitrie Culcer, Yugui Yao, A. H. MacDonald, and Qian Niu}
\affiliation{Department of Physics, The University of Texas at
Austin, Austin TX 78712-1081} \affiliation{International Center
for Quantum Structures, Chinese Academy of Sciences, Beijing
100080, China}
\date{\today}
\begin{abstract}
We propose a simple way of evaluating the bulk spin generation of
an arbitrary crystal with a known band structure in the strong
spin-orbit coupling limit. We show that, in the presence of an
electric field, there exists an intrinsic torque term which gives
rise to a nonzero spin generation rate. Using methods similar to
those of recent experiments which measure spin polarization in
semiconductors, this spin generation rate should be experimentally
observable. The wide applicability of this effect is emphasized by
explicit consideration of a range of examples: bulk wurtzite and
strained zincblende ($n$-GaAs) lattices, as well as quantum well
heterojunction systems.
\end{abstract}
 \maketitle
\section{Introduction}
A considerable amount of attention has been devoted in recent
years, both experimentally and theoretically, to the field of
semiconductor spintronics. There have been numerous suggestions of
advantages offered by the manipulation of the spin degree of
freedom, including the increased functionality of spin devices,
low power consumption, integration with existing technologies and
the fact that in spin transport quantum coherence can be
maintained on large time scales. Semiconductors can be used in the
same spintronic devices as metals, and have the additional
advantages brought about by the existence of an adjustable bandgap
and by the ability to manipulate the carrier density over many
orders of magnitude by doping, gating and heterojunction
formation. Moreover, spin accumulation in a semiconductor will
generate a much larger voltage because the density of states at
the Fermi energy is lower than in a metal, leading to a larger
spin splitting \cite{Fert}.

To this end it would be extremely desirable if a practical method
existed for the efficient generation of spin polarization inside a
semiconductor, as well as for the transport of spins over a
sizable length scale. Although optical spin injection \cite{Young}
has been known for decades, it is impractical for devices, since
it is not sufficiently local for nanoelectronics. On the other
hand, spin injection from a ferromagnetic metal into a
semiconductor requires long spin lifetimes \cite{Kikkawa} and is
impeded by the resistivity mismatch between the two materials
\cite{Schmidt}. Progress has been achieved with ferromagnetic
semiconductors \cite{Schmidt, YOhno, Fiederling, Jiang, Jonker,
Zhu}, but room temperature semiconductor ferromagnetism
\cite{HOhno} has not yet been clearly established. Recent research
has focused on the electrical control of spins in semiconductors
\cite{Prinz, Wolf, Das}, including theoretical work on the
possibility of observing a spin-Hall effect \cite{Culcer, SpinHall},
together with various efforts to generate a spin current
\cite{Spincurrent,AwschSC,Wunderlich}.

Following this energetic theoretical enterprise, recent
experimental work \cite{Awschalom, Hammar} has demonstrated the
detection of a sizable spin accumulation in semiconductors to be a
feasible task. Motivated by these findings, in this article we
present a general theory of the intrinsic electrical spin
generation that occurs generically in spin-orbit coupled systems
that are not inversion-symmetric. We will concentrate on systems
where the spin-orbit interactions are strong. The existence of the
mechanism we discuss has been pointed out in the previous
literature \cite{Levitov, Edelstein, Aronov, Chaplik}. Our
approach is to be contrasted with earlier theories which consider
directly the response of a spin polarization to an electric field,
by Aronov {\it et al.}\cite{Aronov} for example. When
calculating the spin density directly via linear response it is,
in general, not possible to separate the intrinsic generation
terms from the extrinsic scattering effects, thus physical
transparency is often sacrificed. On the other hand, in the
present paper we recognize spin generation as an intrinsic effect
so that it may be determined from first principles calculations.
The interplay of the spin generation and relaxation terms,
resulting in a finite spin polarization, emerges in the final
analysis, as pointed out by Edelstein\cite{Edelstein}. So far,
experiment \cite{Awschalom} has attempted to measure the spin
generation rate separately from the spin relaxation time in the
{\it weak} spin-orbit coupling limit. The idea we discuss has
already been applied to the Rashba Hamiltonian \cite{Chaplik},
which we also examine below.

Spin-orbit interactions can be important in semiconductors for
several reasons. The first is the fact that the carriers are
clustered near the band extrema around high symmetry points where
there exist degeneracies, and the form of these degeneracies is
determined by the spin-orbit interaction. Due to the fact that in
semiconductors the carriers occupy a narrow width of $k$-space the
spin-orbit interaction can therefore play a crucial role. We will
show in this article that, because of the fact that spin is not
conserved, there exists a term which acts as a bulk source of spin
generation. It represents the rate of change of the spin density
in response to an external electric field.

Intrinsic spin-orbit effects have been shown to lead to a non-zero
Berry curvature which gives a contribution to the anomalous Hall
effect \cite{AHE, Tomas, Yao}, while a spin-orbit-induced
metal-insulator transition has been detected by Koga {\it et al.}
\cite{Koga}. Aside from the references mentioned above, such
intrinsic spin-orbit effects have not been taken into account
previously, although several theories exist which account for the
role intrinsic spin-orbit effects play in spin relaxation
\cite{SOscatt}. Moreover, a number of articles have considered an
electric-field-induced rotation of the total spin
polarization\cite{Spingenrel, Drift-diffusion}, which is distinct from the
electrical spin generation we discuss in our work. In the
situations we consider, the total spin polarization is initially
zero, therefore it cannot undergo a rotation.

Because both the rate of change of the spin density and the
electric field are even under time reversal, ferromagnetism is not
required for spin generation. Nevertheless, because the electric
field changes sign under spatial inversion while the rate of
change of the spin density does not, spin generation occurs only
in crystals with broken inversion symmetry. Further details of
these symmetry arguments will be given in the last section.

The outline of this paper is as follows. In section II we will
present the formalism underlying our discussion of intrinsic spin
generation. The mechanism we outline applies to a wide class of
systems, and to illustrate this we discuss, in section III,
two-dimensional heterostructures described by the Rashba model,
followed by a model of the conduction bands of unstrained bulk
wurtzite structures and strained bulk $n$-GaAs.

\section{Formalism}
In the physical picture we consider, the spin-orbit interaction
has been taken into account in the band structure. We adopt a
Boltzmann-wave-packet approach, in which the carriers in a band labeled by the index $n$ are
described by wave-packets following a Boltzmann phase-space
distribution $f_n({\bf r}_c, {\bf k}, t)$. The construction of a
wave-packet representing a charge and spin carrier in band $n$, which has real
and $k$-space coordinates $({\bf r}_c, {\bf k})$, has been
thoroughly treated by Sundaram and Niu\cite{Sundaram} and will not
be considered at length here. The semiclassical equations of
motion for $({\bf r}_c, {\bf k})$, in the presence of a constant
uniform electric field {\bf E},
are\cite{Sundaram}:\begin{equation} \label{sc}\begin{split}\hbar
\dot {\bf k}_c = -e{\bf E}\\ \hbar \dot {\bf r}_c =
\pd{\varepsilon_n}{{\bf k}_c} + e{\bf E} \times {\bf \Omega}_n,
\end{split}\end{equation}
where ${\bf \Omega}_n = 2{\rm Im} \dbkt{\pd{u_n}{{\bf k}}}{\pd{u_n}{{\bf k}}}$ represents the Berry, or geometrical curvature\cite{Sundaram}.

The theory presented in this article is a theory of spin
accumulation in the strong spin-orbit coupling limit, implying
that the splitting of the bands due to the spin-orbit interaction
exceeds their broadening due to disorder at all wave-vectors. As a
result, the bands are well defined although, due to the presence
of the spin-orbit interaction, they are not pure spin-up and
spin-down bands. The distribution function corresponding to each band $n$, $f_n$, can drift according to the semiclassical
equations of motion, and can also change due to collisions. The
time evolution of the distribution function is governed by the
Boltzmann equation, \beq \label{Boltz} \pd{f_n}{t} + \dot{\bf
r}_c\cdot\pd{f_n}{{\bf r}_c} + \dot{\bf k}\cdot\pd{f_n}{{\bf k}} =
\td{f_n}{t}|_{coll}.\eeq The right hand side, $\td{f_n}{t}|_{coll}$,
is the collision term which may be modeled by a relaxation time
approximation when scattering is weak enough, or it may be
expressed in terms of collision integrals. For electrons in
equilibrium, the solution of the Boltzmann equation is $f_n^{(0)}$, the
Fermi-Dirac distribution, while in a general nonequilibrium
situation $f_n$ can be written as $f_n^{(0)} + \delta f_n$. In the presence
of a constant uniform electric field {\bf E} the shift $\delta f_n$
is given, in the relaxation time approximation, by the well-known
result\cite{AM}:
\begin{equation}
\delta f_n = f_n - f_n^{(0)} = e\tau_p{\bf E}\cdot{\bf
v}_n\pd{f_n^{(0)}}{\varepsilon},
\end{equation}
where ${\bf v}_n = \frac{1}{\hbar}\pd{\varepsilon_n}{{\bf k}}$, $\varepsilon_n$ is the energy of band $n$, and
$\tau_p$ is the momentum relaxation time. We shall refer to terms
which depend only on the equilibrium value of the distribution
function, $f_n^{(0)}$, as {\it intrinsic}, as opposed to {\it extrinsic}
terms, depending on the nonequilibrium distribution, and therefore
on scattering.

The study of spin generation necessarily relies on the spin
equation of continuity. For the case we consider, we have shown in
a previous publication\cite{Culcer} that this equation takes the
form: \beq \label{continuity} \pd{S_n}{t} + \nabla\cdot{\bf J}^{s}_n =
{\cal T}_n + \int d^3k\td{f_n}{t} \bkt{\hat s}_n. \eeq The terms on the
LHS represent the spin density and current in band $n$, while the last term in
the equation takes into account collisions. The abbreviation $\bkt{\hat s}_n$ stands for the expectation value in band $n$ of the spin operator corresponding to any one component of the spin. We specialize
henceforth in homogeneous systems, in which the divergence of the
spin current in (\ref{continuity}) will be zero and the equation of continuity may be written as: 
\beq \label{cont2} \pd{S_n}{t} = {\cal T}_n + \int d^3k \frac{(f_n^{(0)} - f_n)}{\tau_p} \bkt{\hat s}_n, \eeq
where $\tau_p$ is the relaxation time. The first term on the RHS, which accounts for spin generation in the absence of collisions, is the focus of this paper. This term, which we shall
call the torque density, exists due to the fact that spin is in
general not conserved and therefore the average spin of a
wave-packet is not constant in time. As discussed in Culcer {\it
et al.}\cite{Culcer}, this torque density is defined as:
\beq\label{firstprinciples} {\cal T}_n({\bf r}, t) \equiv \int d^3
r_c \int d^3k f_n({\bf r}_c, {\bf k}, t)\bkt{\hat {\tau}\delta({\bf
\hat r} - {\bf r})}_n, \eeq in which ${\hat \tau}$ is understood as
$\frac{i}{\hbar}[\hat H, \hat s]$, $\hat H$ is the Hamiltonian,
${\bf \hat r}$ is the quantum-mechanical position operator and $\bkt{}_n$ stands for the expectation value in a wave-packet constructed starting from the eigenfunction corresponding to band $n$. Throughout this paper we assume that products of non-commuting operators have been symmetrized. We
note that the equation of continuity (\ref{continuity}) is derived
directly from the first-principles definitions of $S_n$, ${\bf J}^s_n$
and ${\cal T}_n$\cite{Culcer}. In homogeneous systems the torque
density simplifies to \beq\label{torque} {\cal T}_n = \int d^3k
f_n\bkt{\hat \tau}_n, \eeq which will be referred to as the {\it spin
generation term}. The fact that we are considering homogeneous
systems also implies that we may regard the wave-packets as being
wide in real space, thus sharp in $k$-space, and evaluate the
expectation value $\bkt{{\hat \tau}}_n$ using Bloch wavefunctions.
It should be pointed out that, although we arrive at our results
semiclassically, one does not require a local description to
obtain them, and they can be found using, for example, a Kubo
formula approach. The connection to this latter approach will be
detailed in what follows. Moreover, our theory is not restricted
to the generation of spin, since $\bkt{\hat s}_n$ may represent the
wave-packet expectation value of any component of any other
non-conserved observable.

We will be concerned with a system in which only
a constant uniform electric field is present. Making a convenient
choice of gauge, this electric field can be included in the
Hamiltonian through the electromagnetic vector potential {\bf
A}({\bf r}, t) only. This results in a nonadiabatic mixing of the
bands, so that the Bloch wavefunctions $\ket{u_n}$ have the
following form: \beq \label{uptb} \ket{u_n} =\ket{\phi_n} -
\sum_{m\ne
n}\frac{\tbkt{\phi_m}{i\hbar\td{}{t}}{\phi_n}}{\epsilon_n -
\epsilon_m}\ket{\phi_m}, \eeq where the $\{\ket{\phi_n}\}$ are the
unperturbed Bloch eigenstates. The only time dependence comes from
the fact that {\bf k} drifts under the action of the electric
field, as in (\ref{sc}). Therefore it is legitimate to make the
replacement $\td{}{t} = -\frac{e{\bf E}}{\hbar}\cdot\pd{}{{\bf
k}}$ in (\ref{uptb}). In this way, the wave functions $\ket {u_n}$
depend on the electric field through a reactive term, in other
words the field induces a change in the wave functions at each
{\bf k}. The $\{\ket {u_n}\}$ form a complete set. Moreover, the
expectation value $\bkt{\hat s}_n \equiv \tbkt{u_n}{\hat s}{u_n}$
is a function of {\bf k} only, and its time dependence arises
implicitly through its dependence on the wave vector. We have thus
included the effect of the electric field in mixing the bands but
neglected inter-band scattering.

In the limit of wide wave-packets, it is straightforward to prove,
starting from Eq. (\ref{uptb}), that $\bkt{{\hat \tau}}_n$ evaluated
in the $\{ \ket{u_n} \}$ basis is equal to $\td{\bkt{\hat s}_n}{t} \equiv \dot{\bf k}\cdot\pd{\bkt{\hat s}}{{\bf k}}$,
where $\bkt{{\hat s}}$ is evaluated in the unperturbed
$\{\ket{\phi_n}\}$ basis and $\dot{\bf k} = -\frac{e{\bf E}}{\hbar}$ from (\ref{sc}). The former approach is equivalent to
using the Kubo formula to find the response of $\hat \tau$ to an
electric field. Following this line of thought, we find that the
spin generation term is always at least first order in the
electric field. Then, to first order in the electric field, we may replace $f_n$ by its equilibrium value $f_n^{(0)}$. This spin generation term is then purely
{\it intrinsic}, as defined at the beginning of this section. The
final form of this term is:
\begin{equation}
\begin{split}
{\cal T}_n = -{e{\bf E}\over \hbar}\cdot\int d^3k f_n^{(0)} \pd{\bkt{\hat
s}_n}{{\bf k}}.
\end{split}
\end{equation}
Our theory thus shows that there exists a spin generation rate
which can be interpreted as due to a displacement of the
wave-packet in $k$-space, but also as the expectation value of the
operator $\hat{\tau}$ in a state which is not an eigenstate of the
crystal Hamiltonian. Experiment\cite{Awschalom} has been
attempting to measure this rate of generation, but for the time
being success has been achieved only in the weak spin-orbit
coupling limit\begin{footnote}{The experiment described in Ref.
[18] uses a magnetic field to rotate the spins in the $z$
direction, then measures the accumulation of $s_z$. This procedure
is appropriate in the weak spin-orbit limit. In the limit of
strong spin-orbit interaction, if one attempts to rotate the
spins, one must also take into account the effect of the
spin-orbit interaction on the spins.}\end{footnote}.

In the steady state the equation of continuity becomes simply:
\begin{equation}
\int d^3k \frac{(f_n - f_n^{(0)})}{\tau_p} \bkt{\hat s}_n = {\cal T}_n
\end{equation}
Remembering that $\int d^3k f_n^{(0)} \bkt{\hat s}_n = 0$ and assuming a
momentum relaxation time independent of wave-vector, the LHS is
simply $\frac{S_n}{\tau_p}$. We can then rewrite the above equation
to express the steady state spin density as:
\begin{equation}
S_n = {\cal T}_n\tau_p
\end{equation}
The characteristic time governing the relaxation of the spin
density distribution is the momentum relaxation time. This is to
be expected, since the theory describes non-degenerate, well
defined bands.

We note that the torque term must be present even in a clean
system if the Hamiltonian contains spin-non-conserving terms. In
the presence of scattering mechanisms, the {\it intrinsic} spin
{\it generation} term is balanced by the {\it extrinsic} spin {\it
relaxation} term so that a net spin polarization can be reached in
the steady state. In addition, in the systems we consider, we
assume scattering is strong enough to keep the distribution
function near equilibrium and the scattering time small, but not
strong enough to make inter-band mixing important.
\section{Examples}
In order to clarify the significance of the examples presented
below, we start with some comments regarding the spin-orbit
interaction and asymmetry. The terms in the spin-orbit interaction
which are odd in $k$ rely upon the inversion asymmetry of the
system under study. This asymmetry can be of two kinds, depending
on the dimensionality of the system. In three dimensions, the
inversion asymmetry is a property of the underlying material, and
is referred to as bulk inversion asymmetry (BIA). In two
dimensions, an asymmetric confinement potential can provide an
additional source of inversion asymmetry, known as structure
inversion asymmetry (SIA). Moreover, the application of strain
along a particular direction further reduces the symmetry of the
structure, with important consequences which will be examined
below.
\subsection{Rashba-type spin-orbit interaction}
Our theory allows us to treat bulk semiconductors and quantum
wells on the same footing. We begin with a study of the Rashba
Hamiltonian. This Hamiltonian describes, based on symmetry
arguments, the SIA of quantum well or heterojunction based two
dimensional electron systems and is usually the dominant source of
spin-orbit coupling in these systems. The effective Hamiltonian
has the form: \beq H = \frac{\hbar^2 k^2}{2m} + \alpha({\bf
\sigma}\times{\bf k})\cdot{\bf \hat z}, \eeq in which $\alpha$ is
the spin orbit constant and ${\bf \sigma}$ is the vector of Pauli
spin matrices. $\alpha$ is usually taken to have the form
$a_{46}E_z$, where $a_{46}$ is a material specific parameter while
$E_z$ is the component of the electric field in the $z$ direction.
This electric field is in general a function of position in the
quantum well, and there exists in principle an additional
contribution if the interface on one side of the quantum well is
different from the interface on the other side. This effective
Hamiltonian therefore provides only an approximate description of
the Rashba effect. The magnitude of the Rashba interaction can be
tuned by an external gate voltage by an amount which has been
shown to be as much as 50$\%$\cite{Nitta}. It is customary to view
the term multiplying the spin as a momentum-dependent effective
magnetic field \cite{SpinHall} in which the spin precesses. In the
absence of an external magnetic field, the bands in the Rashba
model are degenerate at ${\bf k} = 0$, and each band contains the
same number of spin up and spin down carriers \cite{Winkler}. The
Hamiltonian has eigenvalues $\epsilon_\pm = \frac{\hbar^2 k^2}{2m}
\pm \alpha k$, which will be labeled by + and - respectively. The Berry curvatures are ${\bf \Omega}_{\pm} = \mp\frac{1}{2}\lim_{H\rightarrow 0}\frac{\alpha^2H\hat{\bf z}}{(\alpha^2k^2 + H^2)^{3/2}}$. The
spin generation term takes the following form: \beq \bkt{{\bf \hat
\tau}}_\pm = \mp \frac{e {\bf k}}{2k^3}({\bf k}\times{\bf
E})\cdot{\bf\hat z}, \label{Rashbatorque} \eeq with $\bkt{{\bf
\hat \tau}}$ defined in Eq.(\ref{torque}). Interestingly, the spin
torque does not depend on the spin-orbit constant. However, the
total torque term, summed over the two bands, depends on the
difference in Fermi wave vectors, which is proportional to the
spin-orbit constant. We find that \beq {\bf \cal T} =
\frac{e\alpha m}{4\pi\hbar^2}{\bf E}\times{\bf \hat z}, \eeq which
vanishes in the limit in which $\alpha \rightarrow 0$, is in
agreement with Magarill\cite{Chaplik}. This result does not depend
on the number density and has a universal form, but it should be
noted that its validity is not universal, rather it is restricted
to systems in which disorder is weak as discussed at the end of
the previous section.

Using symmetry arguments, we find that the spin-orbit coupling in
the conduction band of bulk wurtzite structures is also described
by a Rashba-type Hamiltonian, with a spin-orbit constant defined
analogously. This conclusion is supported by group theory
arguments\cite{Cardona}. The only terms linear in $k$ (and
therefore dominant except at high densities) which are allowed by
symmetry are $\beta(\sigma_x k_y - \sigma_y k_x)$, and the
Hamiltonian is: \beq H=\frac{\hbar^2k^2}{2m} + \beta(\sigma_x k_y
- \sigma_y k_x) \eeq with eigenvalues (labeled as before)
$\frac{\hbar^2k^2}{2m} \pm \beta k$. The spin generation term has
a form very similar to (\ref{Rashbatorque}): \beq \bkt{{\bf \hat
\tau}}_\pm = \mp \frac{e {\bf k}_\perp}{2k_\perp^3}({\bf
k}_\perp\times{\bf E})\cdot{\bf\hat z} \label{Wurtztorque} \eeq In
the above, ${\bf k}_\perp = (k_x, k_y, 0)$. The total torque term
is: \beq \label{Rashba}{\bf \cal T} =
\frac{em\beta}{4\pi^2\hbar^2}(3\pi^2n)^{1/3}{\bf E}\times{\bf \hat
z} \eeq which again vanishes as $\beta \rightarrow 0$ and as the
number density $n\rightarrow 0$.

\subsection{Cubic Dresselhaus spin-orbit interaction}
Finally, we turn our attention to the conduction band of
zincblende semiconductors, which has been the focus of experiment
recently \cite{Awschalom}. In order to be close to experiment, we
consider a degenerate electron gas in an $n$-doped
In$_x$Ga$_{1-x}$As heterostructure grown on GaAs, with x=0.07,
with a strain of 0.46$\%$ directed along (001), as given in the
recent paper of Kato {\it et al.}\cite{Awschalom} and references
therein. The Fermi surface is only slightly displaced from
equilibrium, as revealed by the drift velocities and by the
mobility measurements in the supplementary table. We will not,
however, attempt to simulate experiment, as many aspects seem to
remain incompletely understood. For example, it is known that the
strain tensor acquires off-diagonal shear terms\cite{Kavanagh},
but experiment does not so far provide an unambiguous way of
determining their role\cite{Awschalom, Bernevig}. Therefore, we
will consider an idealized situation in which the $x$ and $y$
lattice constants are equal to their substrate values, while the
lattice constant in the $z$ direction expands according to the
elastic equations. Moreover, there is no convincing pattern in the
variation of the BIA with increasing strain in the experiment. In
fact, the only sizable increase is observed at the largest value
of the strain, which is $0.46 \%$. Thus, it is not clear whether
the linear or cubic term in the BIA is dominant in the samples
investigated.

The symmetry of zincblende does not allow terms linear in $k$ in
the conduction band in the bulk. As a result, when strain is
applied these linear-in-$k$ terms will be first order in the
strain. These terms will be important at small wave vectors, but
we will concentrate on situations in which the number density
makes the Fermi wave-vector $k_F$ high enough that the cubic
Dresselhaus term is dominant. We therefore neglect the effect of
the terms linear in $k$ in this calculation, although we take into
account the effect of strain on the effective masses. We will take
into account only the spin-orbit terms cubic in $k$ which are
present in the unstrained lattice \cite{Winkler}, namely
$\lambda\sigma_x [k_x(k_y^2-k_z^2) + c.p.]$, where $\lambda$ is
the spin-orbit constant and c.p. stands for cubic permutations. To
determine the range of validity of this approach, we estimate the
doping density $n$ at which the $k^3$ term dominates the
linear-in-$k$ BIA and SIA terms. Based on the data in Kato {\it et
al.}\cite{Awschalom}, we estimate that the term cubic in $k$ will
dominate if $n \ge 2.7 \times 10^{16}$ cm$^{-3}$, which puts the
experiment, in which $n = 3 \times 10^{16}$ cm$^{-3}$, narrowly in
the range in which this term is dominant. Our prediction is,
however, consistent with the bulk BIA findings of Kato {\it et
al.}

Our theory is valid in the limit of strong spin-orbit interactions
or weak disorder. For our theory to be valid, the following must
hold:
\begin{equation}
\frac{\hbar}{\tau_p} < \Delta_{so}(k_F),
\end{equation}
where $\tau_p$ is the momentum relaxation time and
$\Delta_{so}(k_F)$ is the spin-orbit splitting at the Fermi wave
vector. From the mobility, $\tau_p$ is estimated at 0.22ps,
yielding $\frac{\hbar}{\tau_p} = 2.9$ meV. In order for the
spin-orbit to overwhelm this, the number density must exceed $3.5
\times 10 ^{18}$ m$^{-3}$, which is a more stringent requirement
than the requirement that the cubic Dresselhaus term exceed the
linear one. In other words, when the system is in the strong
spin-orbit limit it is already in the regime where the $k^3$ term
dominates. This calculation also shows that the experiment of Kato
{\it et al.} lies in the weak spin-orbit coupling limit, outside
the validity limit of our theory.  The experiment is performed in
a doped semiconductor in the extrinsic regime, where extrinsic (as
defined in section II) refers to situations in which scattering
effects are dominant, that is the spin-orbit splitting at $k_F$ is
broadened beyond the point at which the bands overlap. By
increasing $n$ one increases $k_F$ and therefore
$\Delta_{so}(k_F)$, passing into the intrinsic regime.

The Hamiltonian for this system is: \beq H = \frac{\hbar^2
k_\perp^2}{2m_\perp} + \frac{\hbar^2 k_z^2}{2m_z} +
\lambda\sigma_x [k_x(k_y^2-k_z^2) + c.p.], \eeq where ${\bf
k}_\perp$ has been defined above. It has eigenvalues
$\varepsilon_\pm = \frac{\hbar^2 k_\perp^2}{2m_\perp} +
\frac{\hbar^2 k_z^2}{2m_z} \pm \lambda \Delta$ (labeled as
before), with $\Delta$ given by $\sqrt{[k_x^2(k_y^2-k_z^2)^2
+c.p.]}$. The Berry curvatures are given by ${\bf \Omega}_\pm = \pm\frac{(k_x^2 - k_y^2)(k_z^2 - k_x^2)(k_y^2 - k_z^2){\bf k}}{2\Delta^3}$. In this model the $x$-component of spin takes the form:
\beq \bkt{\hat {s}^x}_\pm =\frac{\hbar
k_x(k_y^2-k_z^2)}{2\Delta},\eeq with the other components given by
cubic permutations of this expression. The $x$-component of the
spin generation term is: \beq \label{finaltau}{\bkt{\hat
{\tau}^x}}_\pm = \mp\frac{e\lambda E_x(k_y^2 - k_z^2)(k_y^2k_z^4 +
k_z^2k_y^4 - k_x^4k_y^2 - k_x^4k_z^2)}{2\Delta^3}. \eeq Again the
other components can be found by cubic permutation. Note that, if
strain were absent so that the effective mass would be isotropic,
the band structure and thus the equilibrium distribution would
have cubic symmetry, making ${\bkt{\hat \tau}}$ zero after
integration over $k$. As a result, we expect that in the strong
spin-orbit coupling limit the spin generation will be proportional
to the strain.

To facilitate comparison with experiment, it is more instructive
to examine the {\it electric torque response tensor}, which we
define through the equation ${\cal T}_i = \chi^\tau_{ij} E_j$, and
$\chi^\tau_{xx} = - \chi^\tau_{yy}$ have finite values, the other
components being zero. In contrast to the Rashba model, the
diagonal components of the tensor are finite, whereas the
off-diagonals vanish. To obtain an explicit expression for the
total spin torque, we must integrate over the Fermi surface, which
in this case is ellipsoidal. We use the equilibrium distribution
and the fact that $f(\varepsilon \pm \lambda \Delta)=
f(\varepsilon)\pm \lambda\Delta \pd{f(\varepsilon)}{\varepsilon}
$, with $\varepsilon=\frac{\hbar^2 k_\perp^2}{2m_\perp} +
\frac{\hbar^2 k_z^2}{2m_z}$ and
$-\pd{f(\varepsilon)}{\varepsilon}=\delta(\varepsilon-\varepsilon_F)$.
We write the mass ratio as $\frac{m_z}{m_\perp} = 1 + \gamma$,
where $\gamma$ is a small quantity, and we find that for the
system under study $\gamma = 0.023$ \cite{Vurgaftman}. The total
result is: \beq\label{etrt} \chi^\tau_{xx} = \frac{3ne\lambda
m_\perp{\cal I}}{8\pi\hbar^2}(1 + \gamma)^{1/2},\eeq where ${\cal
I}$ is a dimensionless angular integral, which contains the
angular part of (\ref{finaltau}), given by:
\begin{widetext}\beq
{\cal I} = \int_0^\pi d\theta\int_0^{2\pi} d\phi
\sin\theta[\sin^2\theta\sin^2\phi -
(1+\gamma)\cos^2\theta]\times\frac{f(\theta, \phi)}{g(\theta,
\phi)} ,\eeq where the functions $f$ and $g$ are as follows:
\beq\begin{split} f (\theta, \phi)= (1 +
\gamma)^2\cos^4\theta\sin^2\phi +
(1+\gamma)\cos^2\theta\sin^2\theta\sin^4\phi -
\sin^4\theta\cos^4\phi\sin^2\phi -
(1+\gamma)\cos^2\theta\sin^2\theta\cos^4\phi \el g (\theta, \phi)
= \sin^4\theta\cos^2\phi\sin^2\phi
+(1+\gamma)\sin^2\theta\cos^2\theta(\cos^4\phi + \sin^4\phi) +
(1+\gamma)^2\cos^4\theta  -
6(1+\gamma)\cos^2\theta\sin^2\theta\cos^2\phi\sin^2\phi .
\end{split}\eeq
\end{widetext}
Using a Monte Carlo integration method, we find that ${\cal
I}$=-0.03037. Based on symmetry arguments, ${\cal I}$ vanishes
when $\gamma = 0$, i.e. when $m_z = m_\perp$. This can be seen
from Eq. (\ref{finaltau}) by switching $k_y$ and $k_z$. At small
$\gamma$, $0 < \gamma < 0.1$, our calculations show that ${\cal
I}$ is linear in $\gamma$ and is given approximately by ${\cal I}
= - 1.28 \gamma$. We expect that, for $n = 10^{19}$ cm$^{-3}$,
$\chi^\tau_{xx} = -1.15\times 10^{-7}$ c/m$^2$. This result is
four orders of magnitude larger than that observed in the current
experiments in the weak spin-orbit coupling limit, which should
offer an incentive for doping the samples in order to move into
the strong spin-orbit limit. Meanwhile, for the density $n =
3\times 10^{16}$ cm$^{-3}$ used in experiment we find
$\chi^\tau_{xx} = -3.83 \times 10^{-10}$ c/m$^2$, a number that is
one order of magnitude higher than that observed. We stress again
however that the experiment was performed in the weak spin-orbit
coupling limit.
\section{Symmetry considerations}
The form of the electric torque response tensor in different
systems can be understood based on symmetry arguments. In Table I
we have listed a number of possible spatial transformations which
are relevant to our problem as well as the behavior of the
electric field and spin torque under these transformations. The
transformations considered are {\it not} assumed to be symmetry
operations of the materials under study. $I_m$ refers to spatial
inversion along the $m$-axis, that is $m\rightarrow -m$, while
$R_m$ refers to a rotation of arbitrary magnitude {\it
anticlockwise} about the $m$-axis. For simplicity and without loss
of generality we take the electric field to be directed along $x$
and consider the generation of spin along all three Cartesian axes
in response to this electric field.

Referring to the first part of the table, first row, it can be
seen that under spatial inversion in the $x$-direction the
electric field changes sign while $\bkt{\hat \tau}_x$ remains the
same. On the other hand, under the same transformation $\bkt{\hat
\tau}_y$ and $\bkt{\hat \tau}_z$ change sign, behaving in the same
way as the electric field.
\begin{table}[h!b!p!]
\caption{Behavior of the electric field and torque under various
spatial transformations}
\begin{tabular}{lllll}
\hline
Operation \,\,\,\, & $E_x$\,\,\,\,\,\,\,\,  & $\bkt{\hat \tau}_x$  \,\,\,\,\,\,\,\, & $\bkt{\hat \tau}_y$ \,\,\,\,\,\,\,\, & $\bkt{\hat \tau}_z$  \\
\hline
$I_x$ & $-$ & + & $-$ & $-$ \\
$I_y$ & + & $-$ & + & $-$ \\
$I_z$ & + & $-$ & $-$ & + \\
\hline
$R_x^{\pi}$ & + & + & $-$ & $-$ \\
$R_y^{\pi}$ & $-$ & $-$ & + & $-$ \\
$R_z^{\pi}$ & $-$ & $-$ & $-$ & $+$ \\
\hline
$I_xR_x^{\pi/2}$ & $-$ & + & -$\bkt{\hat \tau}_z$ & +$\bkt{\hat \tau}_y$ \\
$I_yR_x^{\pi/2}$ & + & $-$ & -$\bkt{\hat \tau}_z$ & -$\bkt{\hat \tau}_y$ \\
$I_zR_x^{\pi/2}$ & + & $-$ & +$\bkt{\hat \tau}_z$ & +$\bkt{\hat \tau}_y$ \\
\hline
\end{tabular}
\label{table1}
\end{table}
As a result, if spatial inversion along $x$ is a symmetry
operation of the material the torque term $\bkt{\hat \tau}_x$ will
vanish, but no information can be inferred about $\bkt{\hat
\tau}_y$ and $\bkt{\hat \tau}_z$. An analogous argument holds for
inversion along the $y$ and $z$ axes. Therefore if spatial
inversion along all three axes is a symmetry of the material all
components of the torque must vanish. This is consistent with the
observation that $\bkt{\hat \tau} = 0$ under a three dimensional
inversion. Furthermore, this table helps explain the form of the
torque in systems with Rashba spin-orbit interaction. The Rashba
Hamiltonian has $I_x$ and $I_y$ as symmetries while $I_z$ is
broken. Examining Table I we see that in response to $E_x$ only
$\bkt{\hat \tau}_y$ can be nonzero, consistent with our findings
in the previous section.

Next let us consider the effect of rotations by $\pi$ about the
three Cartesian axes. In the second part of the table, first row,
we examine a rotation about $x$. This rotation does not affect
$E_x$ and $\bkt{\hat \tau}_x$, but $\bkt{\hat \tau}_y$ and
$\bkt{\hat \tau}_z$ change sign. Therefore, if $R_x^{\pi}$ is a
symmetry operation $\bkt{\hat \tau}_y$ and $\bkt{\hat \tau}_z$
will be zero in response to $E_x$. By extension, if rotations by
$\pi$ about all Cartesian axes are symmetry operations then all
the off-diagonal components of the electric torque response tensor
are zero. For example, in GaAs, with or without strain applied
along the $z$-axis, rotations by $\pi$ about all three axes are
symmetry operations so the off diagonal components of $\chi^\tau$
are zero.

Let us also consider the combined effect of rotation and
inversion, taking GaAs as an example. GaAs does not have inversion
along any of the three Cartesian axes as a symmetry operation.
Therefore we cannot infer the vanishing or survival of the
electric torque response tensor in GaAs based on considerations of
inversion alone. With the electric field still along $x$, we
consider an {\it anticlockwise} rotation by $\frac{\pi}{2}$ about
the $x$-axis followed by inversion along $x$, $y$ and $z$. The
rotation does not affect $E_x$ and $\bkt{\hat \tau}_x$. Looking at
the third part of the table, first row, the electric field changes
sign under $I_xR_x$ while $\bkt{\hat \tau}_x$ does not. Therefore,
if $I_xR_x^{\pi/2}$ is a symmetry of the system, the $x$-component
of the torque in response to $E_x$ will vanish. A similar argument
holds for the $y$ and $z$ axes. Therefore if $I_mR_m^{\pi/2}$ is a
symmetry for all Cartesian axes all the diagonal components of the
electric torque response tensor will be zero. This operation is a
symmetry of bulk GaAs and in order to remove the rotational
symmetry we consider strain applied along the $z$ axis. In this
case $I_xR_x^{\pi/2}$ and $I_yR_y^{\pi/2}$ are not symmetries
anymore, allowing $\chi^\tau_{xx}$ and $\chi^\tau_{yy}$ to be
nonzero, while $\chi^\tau_{zz}$ is zero.

\section{Conclusions}
We have shown that in crystals with inversion asymmetry and strong
spin-orbit interactions a spin accumulation will be generated in
the presence of an electric field due to an intrinsic spin torque
term. In the steady state, the spin accumulation is given simply
by the product of this torque term with the momentum relaxation
time. This term is expected to be observable both in systems with
Rashba spin-orbit interactions and with Dresselhaus interactions.
In the latter we have shown that by doping so as to bring the
system into the intrinsic regime, the effect can be observed using
currently available technology.

Finally, we would like to mention a novel experimental method
which has been recently proposed. In condensed matter it is
usually not possible to isolate a single charge and spin carrier
in order to measure effects occurring on the scale of an
individual wave-packet. Therefore, to be able to verify the
existence and properties of the spin generation term one may
resort to atomic physics. Using a cold-atom system to construct an
individual wave-packet and mimic the spin-orbit interaction, an
experiment can measure, for example, the size, wave vector and
electric field dependence of $\bkt{{\hat \tau}}$ for this
wave-packet. This method and the physics underlying it are
described in detail by Dudarev {\it et al.} \cite{Artem}.

DC was supported by the NSF under grant number DMR-0404252. YY
acknowledges the support of the National Science Foundation of
China under grant number 10404035 and by the NSF under grant number DMR-0071893. QN and AHM were supported by
the DOE under grant number DE-FG03-02ER45958.

\end{document}